\documentclass[twocolumn]{aastex701}

\usepackage{graphicx}
\usepackage{amsmath,amssymb}
\usepackage{booktabs}
\usepackage{longtable}
\usepackage{multirow}
\shorttitle{SPHEREx Galactic Backgrounds}
\shortauthors{Lim et al.}
\submitjournal{ApJ}
\received{2026 July 14}
\revised{2026 October 1}
\accepted{2026 October 3}

\begin{document}

\title{Diffuse Galactic Backgrounds with SPHEREx: 
A Pilot Study of Background Modeling and Its Impact on Extinction}

\correspondingauthor{Wanggi Lim}

\author[0000-0003-4243-6809]{Wanggi Lim}
\affiliation{IPAC, California Institute of Technology, 1200 East California Boulevard, Pasadena, CA 91125, USA}
\email[show]{wlim@ipac.caltech.edu}

\author[0000-0002-3470-2954]{Jae Hwan Kang}
\email{jkang7@caltech.edu}
\affiliation{Division of Physics, Mathematics and Astronomy, California Institute of Technology, Pasadena, CA 91125, USA}

\author[0000-0001-6505-0293]{Keunho J. Kim}
\email{keunho11@caltech.edu}
\affiliation{IPAC, California Institute of Technology, 1200 East California Boulevard, Pasadena, CA 91125, USA}

\author[0009-0004-7886-9029]{Young-Jun Kim}
\email{yj\_kim1012@snu.ac.kr}
\affiliation{Department of Physics and Astronomy, Seoul National University, 1 Gwanak-ro, Gwanak-gu, Seoul 08826, Republic of Korea}

\begin{abstract}
We present a spectral analysis of diffuse Galactic background emission using SPHEREx data in a pilot Galactic-plane field, where uncertainty in the estimation of the diffuse infrared background is a major source of systematic uncertainty in extinction studies at near- and mid-infrared wavelengths. We construct channel-resolved background maps across the 0.75--5~$\mu$m range using a large median filter approach and then carry the extinction analysis forward only for the cleaner long-wavelength subset at $\lambda \geq 3~\mu$m. In the present field, comparison against an earlier masked-median background model gives a median channel-by-channel background difference of about 6\% and a maximum change of $|\Delta\tau_{\rm norm}|\simeq 7.4$ near 3.205~$\mu$m, underscoring the importance of wavelength-dependent background modeling. These results demonstrate that spectral mapping with SPHEREx provides a practical framework for controlling diffuse background systematics and enables improved constraints on dust opacity and extinction in the Galactic plane.
\end{abstract}


\section{Introduction}
Diffuse infrared emission dominates the surface-brightness budget over large areas of the Galactic plane and cannot be treated as a small correction in low-latitude analyses \citep{draine03,draine11}. Along these sight lines the observed sky reflects long path lengths through the interstellar medium and the superposition of physically distinct environments, from diffuse atomic gas to dense star-forming structures. The resulting emission field is structured on angular scales from arcminutes to degrees, and its morphology depends simultaneously on dust column, grain properties, illumination geometry, and the mix of emitting phases along the line of sight \citep{draine03,draine11,brandt12,hensley21,chellew22}. In practice, the diffuse background is therefore both an astrophysical observable and a major source of systematic uncertainty when extracting extinction or opacity information from complex Galactic fields. Large-area Herschel Space Observatory (Herschel) and Planck studies have made the same point from the emission side by showing that Galactic dust temperature, emissivity, and column structure vary coherently across arcminute-to-degree scales, so that background estimation is a community-wide problem rather than a peculiarity of infrared-dark cloud (IRDC) analyses alone \citep{bernard10,juvela12,planck14,fanciullo15}.

The physical origin of that background is intrinsically multicomponent. At shorter wavelengths, scattered starlight can contribute substantially, with both amplitude and color controlled by the radiation field, dust properties, and viewing geometry \citep{sellgren84,brandt12,chellew22}. Across the mid-infrared, continuum emission from very small grains and stochastically heated particles becomes increasingly important, while broad polycyclic aromatic hydrocarbon (PAH)-sensitive structure and line emission may dominate in irradiated material \citep{allamandola89,li01,peeters02,dl07,smith07,tielens08,tielens21,zhang25}. These components do not share a common spatial morphology or a common spectral dependence. A small set of broadband images can therefore provide a useful phenomenological description of a field, but not a physically adequate representation of the diffuse background for precision extinction or dust-opacity work.

This limitation propagates directly into several active observational problems. Extinction mapping in dense regions depends on the background intensity behind the absorbing structure, not merely on the observed surface brightness in its immediate vicinity. That dependence enters both near-infrared column-density methods such as NICE, NICER, and NICEST \citep{lada94,lombardi01,lombardi09} and the broader observational literature on optical to infrared extinction curves and their environmental variation \citep{indebetouw05,flaherty07,chapman09,ascenso13,wang14,schlafly16,gordon21}, as well as mid-infrared mass--surface density analyses based on structured diffuse backgrounds \citep{butler09,butler12,battersby11,lim14,lim15,lim16}. The issue is particularly important for IRDCs, i.e., cold dense clouds identified in silhouette against the bright mid-infrared Galactic background in the original Midcourse Space Experiment (MSX)-based discovery and follow-up work of \citet{egan98} and \citet{carey98,carey00}. IRDCs are frequently discussed as candidate initial conditions for massive star and star-cluster formation \citep{mckee03,tan14}, and extinction-based mass measurements have therefore played a central observational role in their study \citep{butler09,butler12}. Compact-source photometry in the inner plane is likewise biased by structured diffuse emission, and the same issue limits the interpretation of bright Galactic fields for facilities such as the James Webb Space Telescope (JWST; \citealt{gardner06}) and future wide-field infrared missions. Earlier Galactic studies necessarily relied on sparse broadband data, local interpolation, or large-scale filtering to estimate the background field \citep{lada94,lombardi01,indebetouw05,flaherty07,butler09,butler12,chapman09,lombardi09,lim14,lim15,lim16}. Those approaches established both the utility and the fragility of background-based extinction measurements: the derived opacity curve is often limited less by formal image noise than by the fidelity of the adopted background estimate itself \citep{butler09,butler12,lim14,lim15,lim16}. A spectrally resolved infrared background framework is therefore relevant not only to opacity-curve work itself, but also to the broader problem of connecting infrared mass estimates, extinction-based column densities, and dust-evolution diagnostics across the Galactic plane.

SPHEREx changes the observational footing of this problem. As an all-sky spectral survey spanning 0.75--5.0~$\mu$m with 102 channels and spectral resolution $R\sim35$--130, it provides dense wavelength sampling across the regime where scattered light, continuum emission, PAH-sensitive structure, and attenuation all change in relative importance \citep{korngut2018,crill20,bock26}. The mission and its sky-simulation framework were designed precisely to support spectrally resolved analyses over wide areas of sky \citep{korngut2018,crill20,crill25,bock26}, while early SPHEREx science studies already demonstrate the diagnostic value of narrow-channel infrared measurements of dusty environments \citep{zhang25}. For the specific problem addressed here, the key gain is not wavelength coverage, but the ability to estimate the diffuse Galactic background channel by channel and then test whether the resulting spectral continuity is sufficient to support a field-specific dust-curve experiment.

The immediate obstacle is technical rather than conceptual. The currently accessible Quick Release 2 (QR2) products are detector-level calibrated spectral images rather than precomputed narrowband diffuse maps \citep{bock26}. That situation differs sharply from the broadband imaging context of the Spitzer Space Telescope (Spitzer; \citealt{werner04}) and Wide-field Infrared Survey Explorer (WISE; \citealt{wright10}), where calibration, atlas production, and coadding strategies have long been documented for community use \citep{reach05,engelbracht07,lang14,mainzer11}. A detector frame may overlap a target field while contributing scientifically useful signal to only a restricted subset of pixels in a requested channel, because the wavelength solution varies across the detector. Channel-resolved background estimation is therefore not just an archive-footprint problem: it is a detector-pixel support problem that must be handled explicitly before any extinction or opacity interpretation is attempted. In this paper we apply that logic to a pilot field near $l\simeq300^\circ$, where the current mosaic set provides uniformly high support across all 102 channels. Our goal is to establish a field-specific framework for SPHEREx diffuse-background estimation and relative-opacity analysis using only the local SPHEREx products, without importing previously published observational $\kappa$ curves from unrelated regions.

\begin{figure*}[t!]
    \centering
    \includegraphics[width=\textwidth]{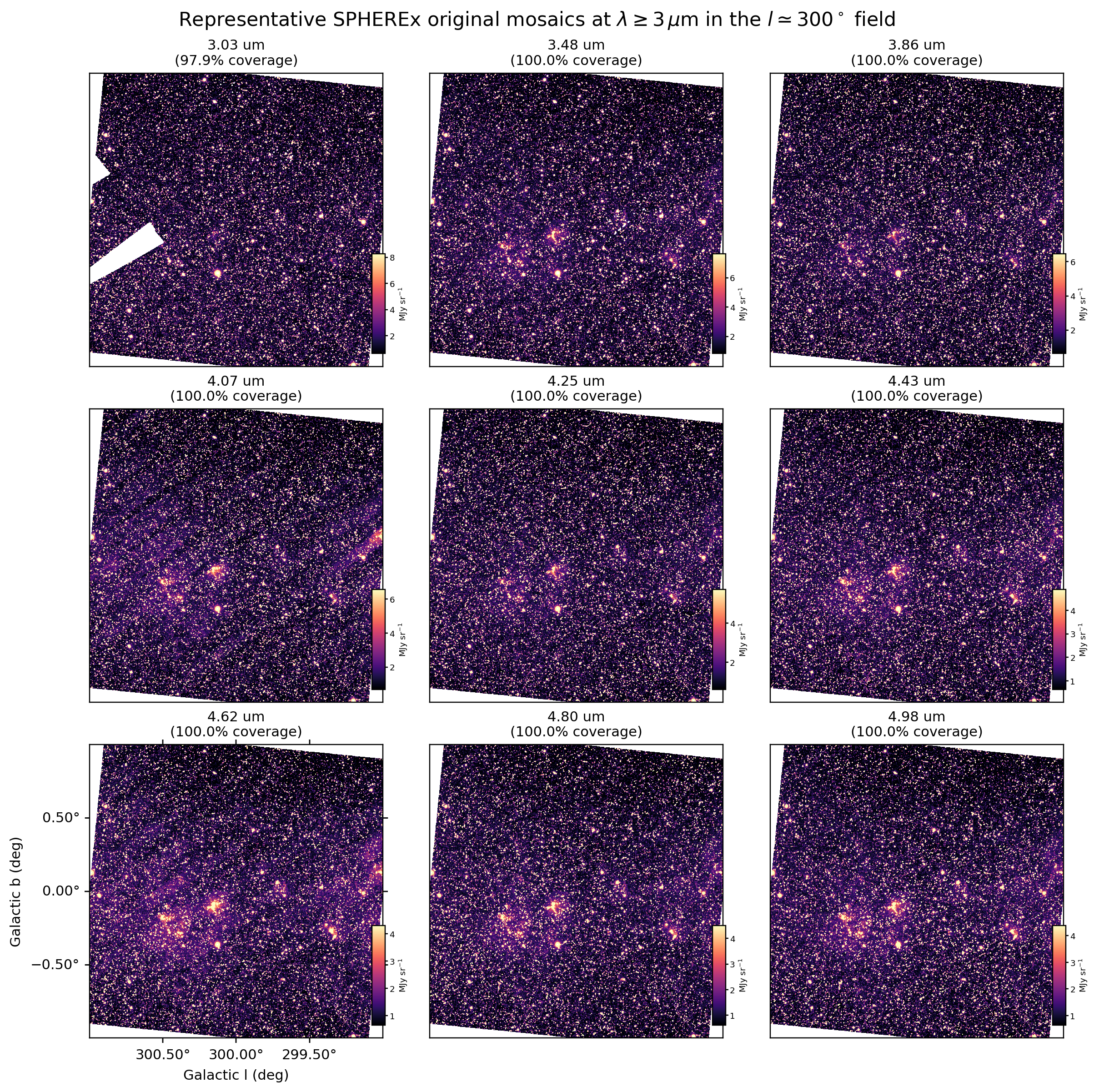}
    \caption{Representative original SPHEREx mosaics in the pilot field centered at $(l,b)=(300^\circ,0^\circ)$. Nine channels are displayed for visual clarity, sampled from the full 102-channel channel set on the common $2^\circ \times 2^\circ$ grid. The figure emphasizes the cleaner long-wavelength part of the sequence used for the pilot extinction and relative-opacity analysis, while the full channel set supports background modeling over all 102 channels.}
    \label{fig:origmulti}
\end{figure*}

\begin{figure*}[t!]
    \centering
    \includegraphics[width=\textwidth]{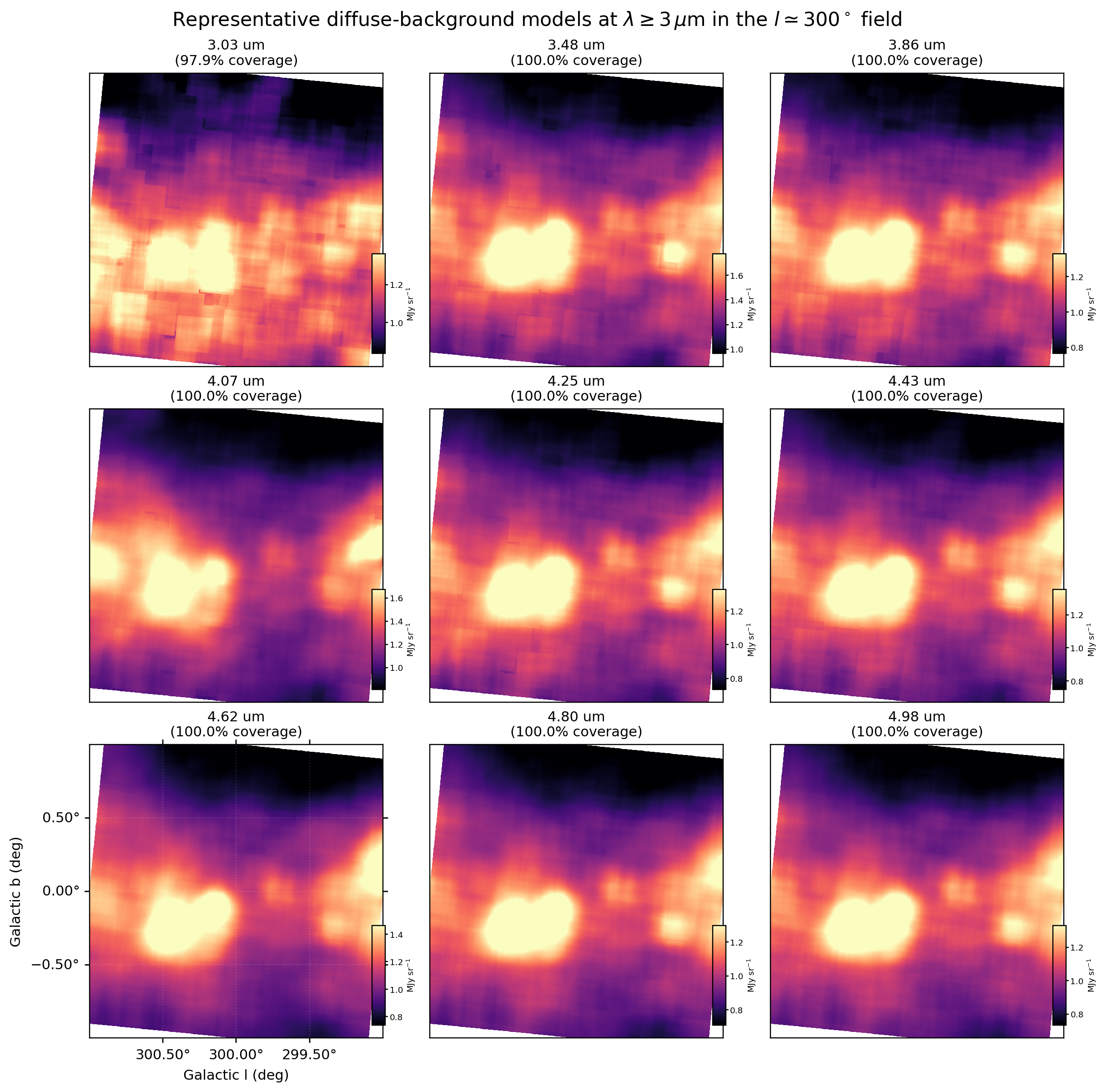}
    \caption{Diffuse-background estimates derived from the same nine representative channels shown in Figure~\ref{fig:origmulti}. These panels illustrate the long-wavelength subset used for the pilot curve, but the 14~arcmin LMF reconstruction of \citet{butler09,butler12} is applied to the full 102-channel set in the current analysis. The retained large-scale coherence across the displayed panels is the basis for the subsequent aperture-based extinction analysis.}
    \label{fig:bgmulti}
\end{figure*}

\section{Data and Methods}

\subsection{SPHEREx Mosaics in the $l\simeq300^\circ$ Field}
We use SPHEREx mosaics generated for this analysis for a Galactic-plane field centered near $(l,b)=(300^\circ,0^\circ)$. The mosaics are assembled from QR2 detector-level products \citep{spherexqr2} and include, for each channel, an intensity image, a zodiacal-light estimate, a zodiacal-subtracted science image, and a hit map. This structure is consistent with the current public SPHEREx mosaic service description, in which a zodiacal-light model is subtracted before mosaic delivery and the NHITS plane records the number of contributing input pixels in each mosaic pixel. We therefore use the hit map as a direct measure of detector-pixel support after channel selection and reprojection, rather than as a binary mask imposed by an archive-side cutoff.

For this field, we generated the full 102-channel SPHEREx mosaic sequence on a common $2^\circ \times 2^\circ$ footprint with 6.15~arcsec pixels. At the time of file retrieval in summer of 2026, this region was among the innermost Galactic-plane fields for which the available QR2 products appeared to provide broad multichannel coverage over the full SPHEREx wavelength range. It therefore provides a useful pilot case for diffuse-background modeling before applying the method to a larger survey area. The current SPHEREx mosaic service documentation also notes that strong default flagging can appear on many sources between 1.6 and 2.4~$\mu$m, often because of the PERSIST flag in the spectral images, and that mosaic pixels can still retain a flux value even when NHITS = 1. We therefore do not impose any additional NHITS cutoff beyond finite-valued support, and we use the hit-map analysis described below only to document the channel-level support across the field rather than to enforce a hard archive-validity threshold.

\begin{deluxetable*}{lc}
\tablecaption{Coverage summary for the SPHEREx pilot field centered at $(l,b)=(300^\circ,0^\circ)$.\label{tab:coverage}}
\tablehead{\colhead{Quantity} & \colhead{Value}}
\startdata
Total mosaicked channels & 102 \\
Supported wavelength interval & 0.75--4.98~$\mu$m \\
All-channel coverage floor & $>96\%$ \\
Minimum channel coverage & 96.9\% \\
Median channel coverage & 99.99\% \\
Channels used for extinction analysis ($\lambda > 3~\mu$m) & 41 \\
\enddata
\end{deluxetable*}

\subsection{Channel Coverage}
For each mosaic, we define the geometric coverage fraction directly from the hit map,
\begin{equation}
C_{\rm ch}=\frac{N\left[H_{\rm ch}(u,v)>0\right]}{N_{\rm tot}},
\end{equation}
where $H_{\rm ch}$ is the hit map for a given channel, $(u,v)$ denotes the pixel coordinates on the output mosaic grid, and $N_{\rm tot}$ is the total number of pixels in the output field. This metric is appropriate for the present pilot study because the extinction analysis requires a common region with genuine data support, not merely a nominal archive footprint.

The full 102-channel set remains uniformly supported. We therefore construct background maps for the full sequence and restrict the opacity analysis only to the cleaner long-wavelength subset described below.

\subsection{Diffuse-background Reconstruction}
For each channel, the zodiacal-subtracted image is used as the input to the background reconstruction. We estimate the diffuse background with the large median filter (LMF) method introduced by \citet{butler09,butler12}. Here the output grid is the common SPHEREx mosaic grid on which each channel image is sampled. A 14~arcmin median window is evaluated on that grid, so the method is defined by an angular scale rather than by a physical grid of $5\times5$ tiles. At an illustrative distance of 5~kpc, this window spans approximately 20~pc, corresponding to a scale characteristic of giant molecular clouds. A smaller window can include more of the absorbing structure and bias the estimated background low, whereas a larger window can miss local background variations. We have not quantified the sensitivity to window size in this study. The coarse implementation is used only for computational efficiency; the $5\times5$ sampling is a numerical shortcut and does not set the physical smoothing scale, which is determined by the 14~arcmin window.

The median calculation is restricted to positive finite pixels, and bright residuals are suppressed by clipping values above twice a robust local mode estimate. The resulting model is defined only where the large-window median can be measured from supported pixels. It is not extrapolated into unsupported regions and is not intended to reproduce compact sources. For the present application, the LMF product is used as a survey-ready estimate of the diffuse emission behind the absorbing structure, while compact-source residuals are treated as contamination for the extinction measurement.

We also measure empirical empty-aperture depths across the same $2^\circ \times 2^\circ$ field as a compact diagnostic of the residual background floor after LMF subtraction. In the Galactic plane, these values are governed mainly by residual diffuse structure and confusion rather than by instrumental variance alone.

\subsection{Extinction and Relative-opacity Proxies}
For the cataloged Spitzer dark cloud SDC300.257$-$0.253 from the \citet{peretto09} and \citet{peretto16} IRDC catalogs, we define an effective optical-depth proxy over a fixed source aperture. We use this source as a pilot aperture, not as a claim that it is the optimal object for a final extinction law. It was selected because it lies in the well-covered $l\simeq300^\circ$ field, remains resolved over multiple SPHEREx pixels, and is surrounded by sufficiently smooth diffuse emission for a first aperture-based test. IRDCs are also relevant pilot targets because they are commonly studied as candidate initial conditions for massive star and cluster formation \citep{mckee03,tan14}.

For that aperture, we measure
\begin{equation}
\tau_{\rm eff}(\lambda)=-\ln\left(\frac{I_{\rm obs}(\lambda)}{I_{\rm bg}(\lambda)}\right),
\end{equation}
where $I_{\rm obs}$ is the observed zodiacal-subtracted intensity and $I_{\rm bg}$ is the reconstructed diffuse background behind the same mask. Foreground emission can, in principle, be constrained using the extinction saturation criterion described by \citet{butler12}. We do not apply that correction here because it requires a sufficiently high column density, an angular extent larger than several beam sizes, and a source geometry that is not strongly biased by nearby emission structures. The purpose of this paper is instead to establish a survey-oriented background modeling framework. The measured $\tau_{\rm eff}$ is therefore used only as a pilot extinction proxy.

For the same spatial structure, we define a relative-opacity proxy by assuming that the column through the selected aperture is fixed with wavelength. The opacity analysis is restricted to 41 channels spanning 3.205--4.98~$\mu$m. This is a methodological science cut rather than a support-driven exclusion: it focuses on the mid-infrared regime where comparison with Infrared Array Camera (IRAC), WISE, Multiband Imaging Photometer (MIPS), Herschel, empirical extinction constraints, and dust models is most informative \citep{indebetouw05,flaherty07,chapman09,ascenso13,wang14,gordon21}. Under the fixed-column assumption
\begin{equation}
\kappa_{\rm rel}(\lambda) \propto \tau_{\rm eff}(\lambda),
\end{equation}
so the normalized opacity proxy has the same spectral shape as the normalized extinction proxy.

In this paper we use two levels of model comparison. For the SPHEREx-only curve in Section~\ref{sec:pilotcurve}, we display the dense-cloud dust models of \citet[hereafter OH94]{ossenkopf94}. These include bare-grain, thin ice-mantle, and thick ice-mantle cases, which provide a compact reference sequence from unmantled grains to increasingly ice-rich and coagulated dust. For the extended comparison in Section~\ref{sec:external} that combines SPHEREx with IRAC, WISE, MIPS, and Herschel, we broaden the model set to include the coagulation models of \citet[hereafter O11]{ormel11}, the laboratory-informed comparison set of \citet[hereafter D22]{dartois22}, and customized dust models constructed with the publicly available modeling tool OpTool \citep{dominik21}. We also evaluate this full set against the SPHEREx-only data to make the effect of adding the external baseline explicit.

The O11 models describe the growth of porous aggregates made from a 2:1 mass mixture of silicate and graphite grains, with a specific ice-coated branch built from ice-coated silicate and ice-coated graphite monomers in vacuum and evolved at $n_{\rm gas}=10^{5}~{\rm cm^{-3}}$ (O11). In the notation used here, the labels O11 1opc through O11 6opc denote the six tabulated opacity snapshots from that adopted ice-coated O11 sequence, corresponding to the published time slices from $3\times10^{4}$ to $10^{7}$~yr in increasing order. We therefore use ``low-order O11'' only to mean the smaller-index, earlier members of this particular O11 opacity sequence, not a separate physical family outside the model grid. The D22 subset used here consists of the E MRN, E tau1, and E tau3 branches (where MRN denotes a Mathis--Rumpl--Nordsieck size distribution; \citealt{mathis77}), which provide a laboratory-informed comparison axis for grain-growth and ice-profile evolution within that framework (D22). The OpTool calculations are designed to isolate the effects of grain-size scaling and ice-mantle mass fraction on the opacity while holding the other dust ingredients fixed. They adopt ``distribution of hollow spheres'' (DHS) grain geometry, with hollow volume fraction = 0.8 and porosity = 25\%, because that setup reproduces the behavior of exact aggregate calculations in a compact parametric form \citep{min05}. The size distribution is MRN-like, with minimum grain size = 0.005~$\mu$m following OH94. The refractory component is a mixture of amorphous pyroxene and amorphous carbon, using the optical constants of \citet{dorschner95} and \citet{zubko96}, with 85\% and 15\% of the refractory mass assigned to silicate and carbon, respectively. The ice mantle uses the H$_2$O:CH$_3$OH:CO:NH$_3$ = 100:10:1:1 optical-constant mixture of \citet{hudgins93}. Within this framework, we vary the maximum grain size from 0.01 to 10~$\mu$m and the ice mass fraction from 0.1 to 0.8, adopting a fiducial model at $a_{\max}=1~\mu$m and ice fraction = 0.2. These extended-comparison models are not treated as a unique physical fit: they are used to illustrate how controlled changes in aggregate state, laboratory-informed ice and grain evolution, grain-size scale, and ice fraction affect the mid-infrared opacity shape within the adopted formalisms. We do not import observational $\kappa$ curves from other sight lines because they are not matched to the present field or to the SDC300.257$-$0.253 aperture.

\begin{table*}[t]
\centering
\caption{Summary of the Dust-model Families used in this work.}\label{tab:dustmodels}
\scriptsize
\setlength{\tabcolsep}{3pt}
\renewcommand{\arraystretch}{1.2}
\begin{tabular}{llllllll}
\hline
\parbox[t]{0.11\textwidth}{\raggedright Model\strut} & \parbox[t]{0.11\textwidth}{\raggedright $n_{\rm H}$\strut} & \parbox[t]{0.08\textwidth}{\raggedright Coag.\strut} & \parbox[t]{0.12\textwidth}{\raggedright Core comp.\strut} & \parbox[t]{0.09\textwidth}{\raggedright Ice mantle\strut} & \parbox[t]{0.13\textwidth}{\raggedright $n(a)$\strut} & \parbox[t]{0.09\textwidth}{\raggedright $a_{\max}$\strut} & \parbox[t]{0.11\textwidth}{\raggedright Main axis\strut} \\
\hline
\parbox[t]{0.11\textwidth}{\raggedright OH94 bare\strut} & \parbox[t]{0.11\textwidth}{\raggedright $10^6\,{\rm cm^{-3}}$\strut} & \parbox[t]{0.08\textwidth}{\raggedright Yes\strut} & \parbox[t]{0.12\textwidth}{\raggedright Silicate + carbon\strut} & \parbox[t]{0.09\textwidth}{\raggedright None\strut} & \parbox[t]{0.13\textwidth}{\raggedright MRN-like\strut} & \parbox[t]{0.09\textwidth}{\raggedright $\sim0.25\,\mu{\rm m}$\strut} & \parbox[t]{0.11\textwidth}{\raggedright Bare-grain limit\strut} \\
\parbox[t]{0.11\textwidth}{\raggedright OH94 thin ice\strut} & \parbox[t]{0.11\textwidth}{\raggedright $10^6\,{\rm cm^{-3}}$\strut} & \parbox[t]{0.08\textwidth}{\raggedright Yes\strut} & \parbox[t]{0.12\textwidth}{\raggedright Silicate + carbon\strut} & \parbox[t]{0.09\textwidth}{\raggedright Thin ice\strut} & \parbox[t]{0.13\textwidth}{\raggedright MRN-like\strut} & \parbox[t]{0.09\textwidth}{\raggedright $\sim0.25\,\mu{\rm m}$\strut} & \parbox[t]{0.11\textwidth}{\raggedright Ice growth\strut} \\
\parbox[t]{0.11\textwidth}{\raggedright OH94 thick ice\strut} & \parbox[t]{0.11\textwidth}{\raggedright $10^6\,{\rm cm^{-3}}$\strut} & \parbox[t]{0.08\textwidth}{\raggedright Yes\strut} & \parbox[t]{0.12\textwidth}{\raggedright Silicate + carbon\strut} & \parbox[t]{0.09\textwidth}{\raggedright Thick ice\strut} & \parbox[t]{0.13\textwidth}{\raggedright MRN-like\strut} & \parbox[t]{0.09\textwidth}{\raggedright $\sim0.25\,\mu{\rm m}$\strut} & \parbox[t]{0.11\textwidth}{\raggedright Ice growth\strut} \\
\parbox[t]{0.11\textwidth}{\raggedright O11 aggregates\strut} & \parbox[t]{0.11\textwidth}{\raggedright Dense cloud\strut} & \parbox[t]{0.08\textwidth}{\raggedright Progressive\strut} & \parbox[t]{0.12\textwidth}{\raggedright Silicate + graphite\strut} & \parbox[t]{0.09\textwidth}{\raggedright Ice-coated\strut} & \parbox[t]{0.13\textwidth}{\raggedright Aggregate\strut} & \parbox[t]{0.09\textwidth}{\raggedright Effective growth\strut} & \parbox[t]{0.11\textwidth}{\raggedright Aggregate growth\strut} \\
\parbox[t]{0.11\textwidth}{\raggedright D22 (E MRN, tau1, tau3)\strut} & \parbox[t]{0.11\textwidth}{\raggedright Dense-core / disk branch\strut} & \parbox[t]{0.08\textwidth}{\raggedright Grain growth\strut} & \parbox[t]{0.12\textwidth}{\raggedright Ellipsoidal silicate\strut} & \parbox[t]{0.09\textwidth}{\raggedright H$_2$O + CO$_2$ ice\strut} & \parbox[t]{0.13\textwidth}{\raggedright Ellipsoidal families\strut} & \parbox[t]{0.09\textwidth}{\raggedright $\lesssim3\,\mu{\rm m}$\strut} & \parbox[t]{0.11\textwidth}{\raggedright Ice-profile distortion\strut} \\
\parbox[t]{0.11\textwidth}{\raggedright OpTool $a_{\max}$\strut} & \parbox[t]{0.11\textwidth}{\raggedright \nodata\strut} & \parbox[t]{0.08\textwidth}{\raggedright No clock\strut} & \parbox[t]{0.12\textwidth}{\raggedright Pyroxene + C (85/15)\strut} & \parbox[t]{0.09\textwidth}{\raggedright Mixed ice, 20\%\strut} & \parbox[t]{0.13\textwidth}{\raggedright MRN, $a_{\min}=0.005$\strut} & \parbox[t]{0.09\textwidth}{\raggedright $0.01$--$10\,\mu{\rm m}$\strut} & \parbox[t]{0.11\textwidth}{\raggedright Size growth\strut} \\
\parbox[t]{0.11\textwidth}{\raggedright OpTool ice frac.\strut} & \parbox[t]{0.11\textwidth}{\raggedright \nodata\strut} & \parbox[t]{0.08\textwidth}{\raggedright No clock\strut} & \parbox[t]{0.12\textwidth}{\raggedright Pyroxene + C (85/15)\strut} & \parbox[t]{0.09\textwidth}{\raggedright Mixed ice, 10--80\%\strut} & \parbox[t]{0.13\textwidth}{\raggedright MRN, $a_{\min}=0.005$\strut} & \parbox[t]{0.09\textwidth}{\raggedright $1\,\mu{\rm m}$\strut} & \parbox[t]{0.11\textwidth}{\raggedright Mantle fraction\strut} \\
\hline
\end{tabular}
\par\vspace{4pt}
\begin{minipage}{\textwidth}\scriptsize
\textit{Note.} Here $n(a)$ denotes the grain-size distribution functional form, not a gas number density. The O11 branch used here is the published ice-coated aggregate sequence built from a 2:1 silicate:graphite mixture at $n_{\rm gas}=10^{5}~{\rm cm^{-3}}$; in our notation O11 1opc--6opc correspond to the six tabulated opacity snapshots from $3\times10^{4}$ to $10^{7}$~yr in increasing order. The laboratory-informed grain-growth models of D22 are considered; the subset used here is built from ellipsoidal silicate grains with H$_2$O and CO$_2$ ice mantles, with some variants also allowing amorphous carbon in the core and CO ice in the mantle. The OpTool models are controlled opacity calculations constructed for this work using the public OpTool framework \citep{dominik21}; in the present model set they adopt DHS geometry with hollow fraction = 0.8 and porosity = 25\%, an MRN-like distribution with $a_{\min}=0.005~\mu$m, refractory material made of amorphous pyroxene and amorphous carbon with an 85:15 silicate:carbon partition, and H$_2$O:CH$_3$OH:CO:NH$_3$ = 100:10:1:1 mantle optical constants.
\end{minipage}
\end{table*}

\subsection{Framework for the External Mid- and Far-infrared Comparison}
The external comparison is performed on the native grids of the archival images. The SDC300.257$-$0.253 aperture defined on the SPHEREx grid is transferred to each external world coordinate system, and the same large-scale background convention is then applied before any extinction proxy is measured.

For the Spitzer/IRAC \citep{fazio04}, we use the native mosaics and the calibration framework of \citet{reach05}. For WISE \citep{wright10}, the atlas cutouts are first mosaicked in native digital-number units. We then construct the LMF background in that native space and calibrate the background-subtracted residual maps to MJy~sr$^{-1}$ units using empirical regressions against matched Spitzer residual maps \citep{jarrett11,jarrett12,jarrett13,mainzer11,lang14}. In the final external comparison we retain only W3 and W4, since the W1 and W2 points are more sensitive to short-wavelength foreground and contrast-suppression systematics in the present pilot field. In the earlier extraction used to assess this cut, the W1 (W2) optical-depth proxies were 0.166 (0.176) versus 0.318 (0.224) for IRAC I1 (I2), which are 48\% (21\%) lower; these adjacent-band discrepancies do not isolate calibration from bandpass and foreground effects. For the Spitzer/MIPS 24~$\mu$m data \citep{rieke04}, four adjacent MIPSGAL tiles \citep{carey09} are mosaicked before background estimation, with the calibration tied to \citet{engelbracht07}. The same prebackground mosaicking step is applied to the two Herschel Photodetector Array Camera and Spectrometer (PACS) 70~$\mu$m blue maps \citep{pilbratt10,poglitsch10}. The resulting extinction proxies are normalized to the native MIPS 24~$\mu$m measurement, keeping the comparison tied to a single field-specific reference without forcing all images onto the SPHEREx grid.

For a quantitative shape comparison, we use
\begin{equation}
\begin{aligned}R_m&=\left[\frac{1}{N}\sum_i(y_i-m_i)^2\right]^{1/2},\\
Q_m&=\frac{1}{N}\sum_i\frac{(y_i-m_i)^2}{s_i^2},\end{aligned}
\label{eq:shape_residual}
\end{equation}
where $i$ indexes the wavelength channels or bands, $y_i$ and $m_i$ are the normalized observed proxy and model opacity, respectively and $y_{16,i}$ and $y_{84,i}$ are the 16th and 84th percentiles of the within-aperture optical-depth distribution, respectively, divided by the same reference optical depth used to normalize $y_i$. We adopt $s_i=(y_{84,i}-y_{16,i})/2$ as a measure of spatial scatter, not as the uncertainty on the representative value. We linearly interpolate the plotted absorption opacities at nominal wavelengths without bandpass integration. The SPHEREx-only and combined tests use $N=40$ and 48, respectively, excluding their respective 3.995~$\mu$m and MIPS normalization points; no parameters are fitted. Because the percentile envelope measures spatial variation and the common normalization correlates points, $Q_m$ is a diagnostic residual, not a reduced $\chi^2$ or a significance measure.

\section{Results}

\subsection{Channel Coverage}
Table~\ref{tab:coverage} summarizes the channel support for the $l\simeq300^\circ$ field. All 102 mosaicked channels remain above 96\% coverage, with a median channel coverage of 99.99\% and a minimum of 96.9\%.

In the present analysis, the hit map simply confirms that the background reconstruction is controlled by sky structure rather than by incomplete support across the full 102-channel sequence.

\subsection{Background Models in the $l\simeq300^\circ$ Field}
Figure~\ref{fig:origmulti} shows nine representative original SPHEREx mosaics sampled from the 102-channel set, with emphasis on the long-wavelength subset used for the dust analysis. Figure~\ref{fig:bgmulti} shows the corresponding LMF backgrounds.

The original images contain substantial compact-source contamination, but broad diffuse structure is visible beyond 3~$\mu$m. After LMF reconstruction, coherent large-scale emission lanes remain identifiable across the displayed sequence. This continuity is the practical reason that the extinction experiment is restricted to the cleaner 41-channel subset spanning 3.205--4.98~$\mu$m, even though background maps are generated for the full 102-channel supported sequence.

An empirical empty-aperture test gives a consistent quantitative summary of that residual background floor: the median background-limited depth is $m_{\rm AB}\simeq12.97$ over all 102 channels and $m_{\rm AB}\simeq13.27$ for the subset at $\lambda\geq3~\mu$m. These are structured-background-limited depths, not formal point-source sensitivities. As the SPHEREx survey continues, both the effective coverage and the background-limited depth should improve as additional visits accumulate. The magnitude of that improvement will vary from field to field, however, because the final gain depends on the local exposure pattern and sky-background structure rather than on elapsed survey time alone.

\subsection{Pilot Extinction and Relative-opacity Curves}\label{sec:pilotcurve}
\begin{figure*}[t!]
    \centering
    \includegraphics[width=\textwidth]{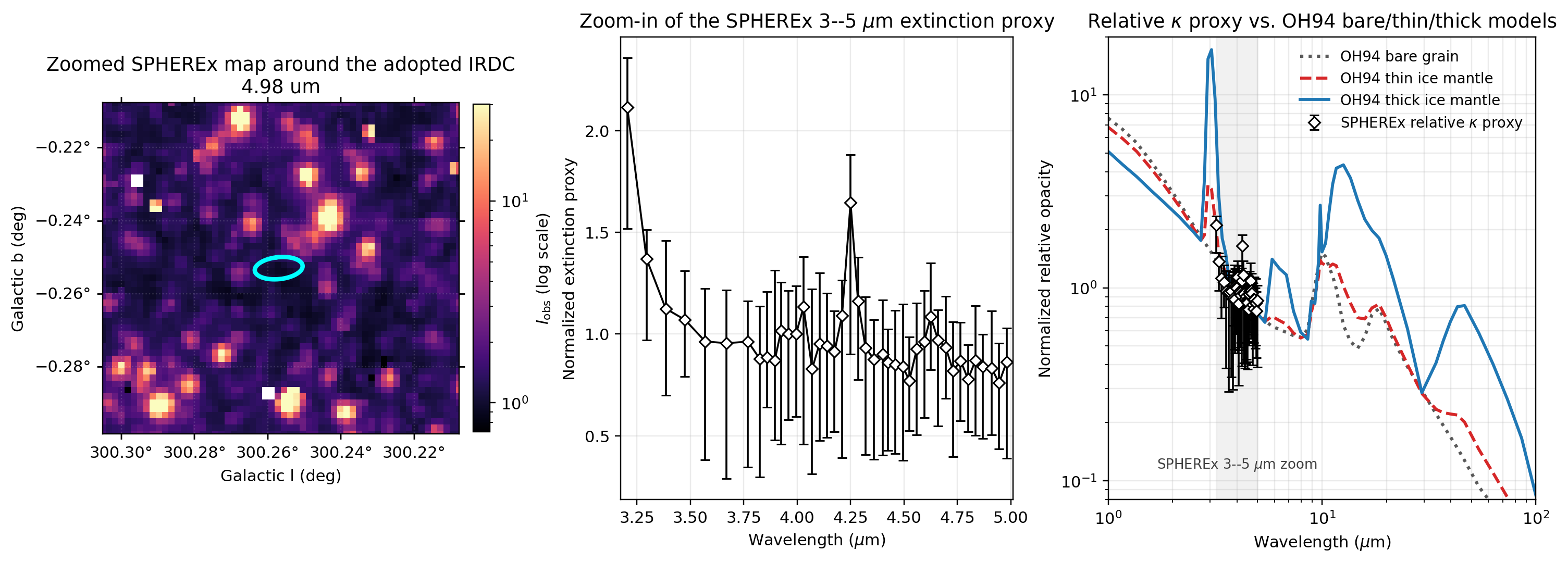}
    \caption{Pilot dust-curve analysis in the field centered at $(l,b)=(300^\circ,0^\circ)$. Left: zoomed 4.98~$\mu$m SPHEREx map around the adopted IRDC SDC300.257$-$0.253, with the aperture highlighted in cyan. Middle: zoomed view of the normalized SPHEREx extinction proxy across the 41 channels spanning 3.205--4.98~$\mu$m used for the long-wavelength science cut. Right: the same normalized spectral shape interpreted as a relative-opacity proxy and compared with OH94 bare-grain, thin ice-mantle, and thick ice-mantle models, normalized at 4.0~$\mu$m; the shaded region marks the wavelength interval enlarged in the middle panel.}
    \label{fig:pilotcurve}
\end{figure*}

Figure~\ref{fig:pilotcurve} presents the pilot dust-curve result for SDC300.257$-$0.253. The left panel shows a zoomed 4.98~$\mu$m SPHEREx view of the adopted IRDC aperture, the middle panel shows a zoomed view of the normalized extinction proxy derived from the 41 channels spanning 3.205--4.98~$\mu$m used in the long-wavelength science cut, and the right panel shows the same spectral shape interpreted as a relative-opacity proxy over the broader comparison baseline.

The curve samples the 3.205--4.98~$\mu$m SPHEREx channel set, a wavelength range where previous extinction studies have found departures from simple extrapolations of optical extinction laws \citep{indebetouw05,flaherty07,chapman09,ascenso13,wang14,gordon21}. The localized enhancements near 3.2 and 4.25~$\mu$m are retained as features of the measured proxy, but they are not interpreted here as secure detections of specific opacity bands. Such an interpretation requires a separate check of $I_{\rm obs}(\lambda)$, $I_{\rm bg}(\lambda)$, and their ratio for the same aperture, since wavelength-dependent background or continuum mismatch can mimic these features.

The first model comparison uses the OH94 models as a compact physical reference set. For the SPHEREx-only data, $(R_m,Q_m)$ is $(0.197,0.251)$ for bare grains, $(0.166,0.189)$ for thin ice, and $(0.888,5.123)$ for thick ice. Thus thin ice has the smallest residual within this family under the stated metric. The lowest-$Q_m$ members of O11, D22, and OpTool give $(0.196,0.241)$ for O11 4opc, $(0.198,0.305)$ for D22 E tau3, and $(0.150,0.147)$ for OpTool $a_{\max}=10~\mu$m, respectively. These are descriptive comparisons of the fixed proxy; they do not measure the statistical significance of a dust-model preference.

The background estimate substantially affects the pilot comparison. A direct comparison between the LMF background and the earlier masked-median estimate gives a median channel-level background difference of about 6\%. Over the 3--5~$\mu$m interval used here, that difference can produce order-unity changes in the normalized extinction proxy, with a maximum change of $|\Delta\tau_{\rm norm}|\simeq7.4$ near 3.205~$\mu$m. Because the background difference is measured in intensity space whereas $\tau_{\rm norm}$ is a normalized logarithmic quantity, modest background offsets can amplify strongly once propagated through the ratio and normalization steps. These differences quantify sensitivity to the adopted background prescription, not its absolute accuracy or the total background uncertainty. Neither method has been independently validated here, and shared biases remain untested. The dust-model comparison is therefore conditional on the adopted LMF background.

\subsection{Extended External Mid- and Far-infrared Comparison}\label{sec:external}
\begin{figure*}[t!]
    \centering
    \includegraphics[width=\textwidth]{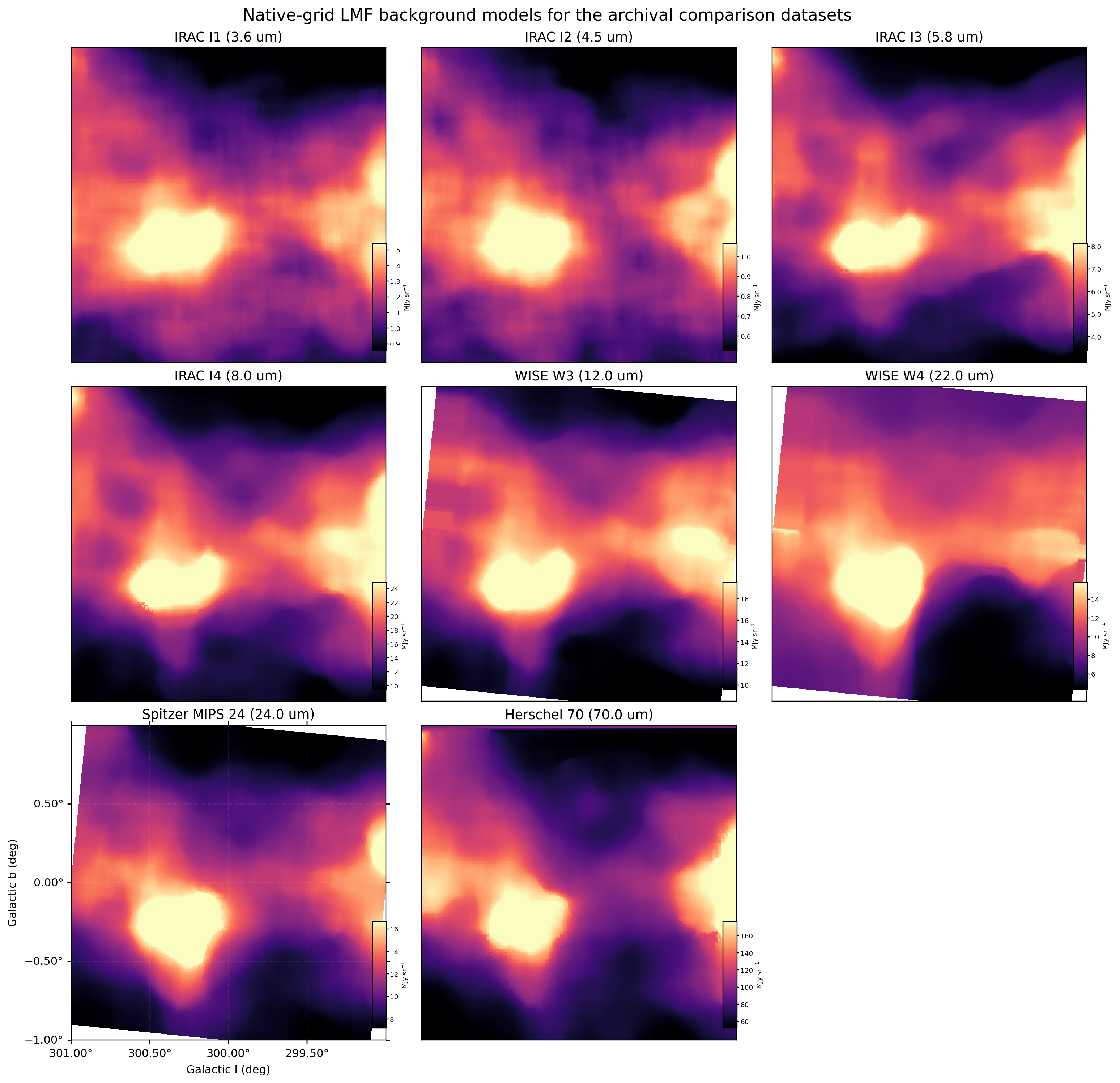}
    \caption{Native-grid LMF background models for the archival comparison datasets, all shown over the same 2$^\circ\times2^\circ$ field in a $3\times3$ layout with the lower-right panel left empty. The four Spitzer/IRAC panels are cropped to the common sky area for direct visual comparison with WISE W3 and W4, MIPS 24~$\mu$m, and Herschel/PACS 70~$\mu$m. The WISE W3 and W4 backgrounds are estimated first in native digital-number units and then transferred to MJy~sr$^{-1}$ units through empirical calibration against matched Spitzer residual maps. The MIPS 24~$\mu$m and Herschel/PACS 70~$\mu$m panels are built from mosaics assembled before background estimation.}
    \label{fig:extbg}
\end{figure*}

The same aperture and background convention are applied to the external IRAC, WISE W3 and W4, MIPS, and Herschel data on their native grids. Figure~\ref{fig:extbg} shows the corresponding 2$^\circ\times2^\circ$ archival LMF backgrounds on a common sky area.

For WISE, the extinction proxy is not taken directly from the atlas surface-brightness scale. Instead, the background is modeled in native digital-number units, subtracted, and then calibrated against matched Spitzer residual maps on coarse angular blocks. The final external-opacity comparison retains only W3 and W4, with W3 tied to an interpolation between IRAC I4 and MIPS 24~$\mu$m and W4 tied to MIPS 24~$\mu$m. This calibration ties the WISE residual structure to the local Spitzer background-subtracted morphology rather than to the atlas zero level alone.

\begin{figure*}[t!]
    \centering
    \includegraphics[width=0.86\textwidth]{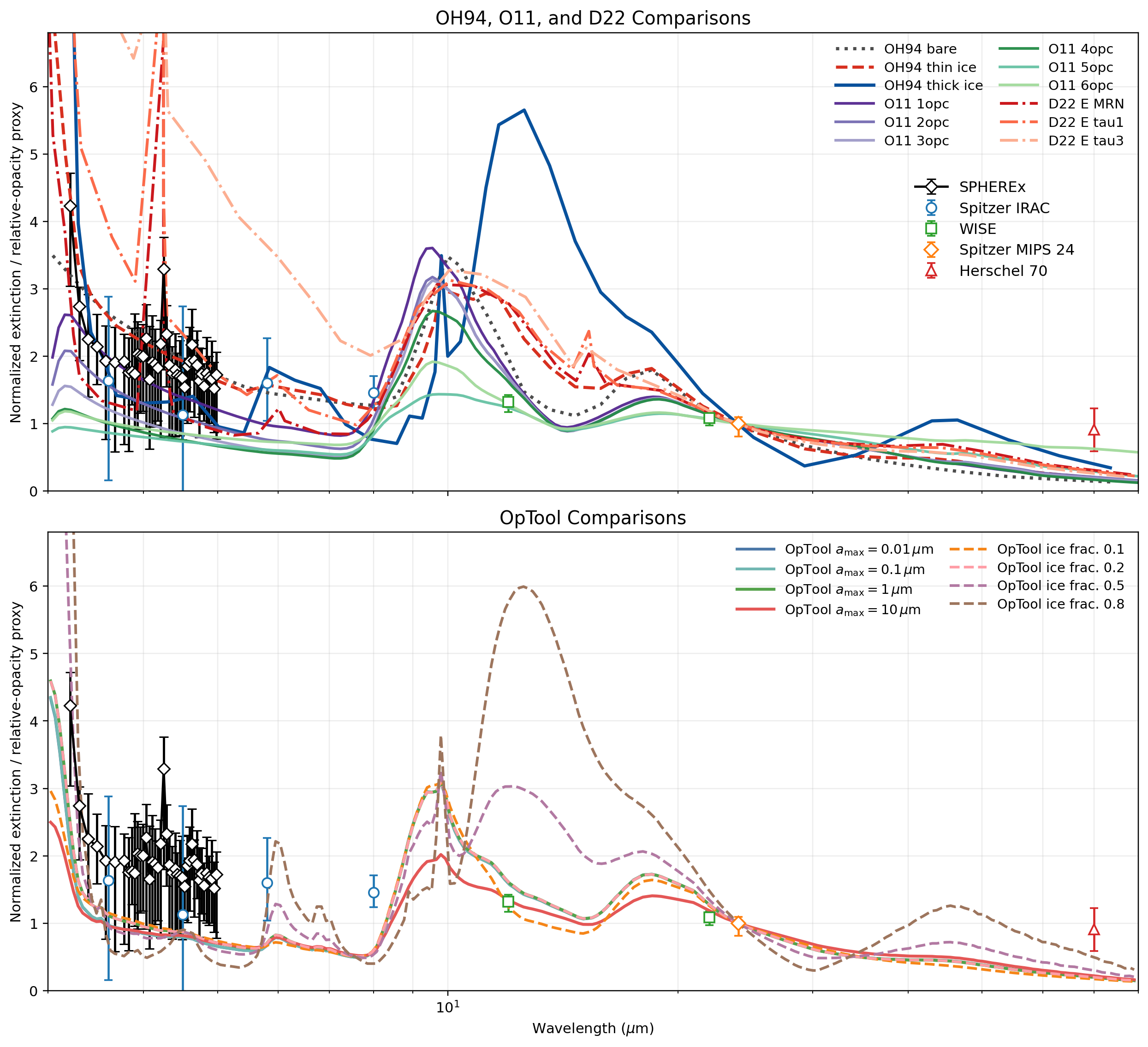}
    \caption{Combined SPHEREx and external archival comparison for SDC300.257$-$0.253 shown in two panels. In both panels, the black curve shows the SPHEREx extinction proxy derived from the 41 channels spanning 3.205--4.98~$\mu$m and renormalized to the MIPS 24~$\mu$m reference point. Blue circles show native-grid IRAC LMF points, green squares show the empirically recalibrated WISE W3 and W4 points, orange diamonds show the MIPS 24~$\mu$m point, and red triangles show the Herschel/PACS 70~$\mu$m point. All external points use the same transferred source aperture, but the comparison is not yet beam matched. Reference FWHM values are approximately 5.6--6.9~arcsec for SPHEREx over the analyzed interval, 1.7--2.0~arcsec for IRAC, 6.5 and 12~arcsec for WISE W3 and W4, respectively, before Atlas smoothing, 6~arcsec for MIPS, and approximately 10~arcsec for PACS 70~$\mu$m in the Hi-GAL maps described by \citet{traficante11} \citep[see also][]{fazio04,rieke04,wright10}. The PACS value is a literature reference, not a measured point-spread function (PSF) for our mosaic. The SPHEREx range is an optical + jitter reference model (\url{https://github.com/SPHEREx/Public-products}), not a measured QR2 mosaic PSF; map processing further modifies the response, so beam dilution can contribute to inter-band offsets. Top: the OH94 bare--thin--thick set, the six O11 ice-coated silicate and graphite aggregate opacity snapshots at $3\times10^{4}$, $10^{5}$, $3\times10^{5}$, $10^{6}$, $3\times10^{6}$, and $10^{7}$~yr (labeled 1opc--6opc in the same order), and the D22 E MRN--tau1--tau3 comparison set. Bottom: the same empirical points compared only against the OpTool realizations.}
    \label{fig:spitzercurve}
\end{figure*}

Figure~\ref{fig:spitzercurve} combines the SPHEREx curve with the external points in two complementary panels. The top panel compares the data with the OH94 models, the O11 coagulation models, and the D22 models, all normalized at 24~$\mu$m. The bottom panel isolates the OpTool-based comparison on the same empirical baseline. The external points therefore provide a broader wavelength baseline without requiring the archival maps to be resampled onto the SPHEREx grid.

The recalibrated WISE W3 and W4 points no longer show the large offset obtained when the atlas surface-brightness scale is used directly. Perfect agreement among the overlapping SPHEREx, Spitzer, and WISE W3 and W4 wavelength ranges is not expected at this stage because the bands sample different bandpasses, native angular responses, zero-level conventions, and local background and foreground residuals. At this stage, the external comparison demonstrates that the LMF background convention can be extended from SPHEREx to archival infrared data for the same source aperture; it does not establish a final extinction law or a unique dust model.

For the combined data, $(R_m,Q_m)$ is $(0.468,1.016)$, $(0.459,2.664)$, and $(1.139,24.319)$ for OH94 bare, thin ice, and thick ice, respectively. The lowest-$Q_m$ members of O11, D22, and OpTool give $(0.529,0.874)$ for O11 1opc, $(1.196,5.674)$ for D22 E MRN, and $(1.012,2.644)$ for the 10\% ice-fraction case, respectively. O11 1opc minimizes $Q_m$, whereas OH94 thin ice minimizes $R_m$; this ordering depends on weighting and does not establish a unique dust population.

\section{Discussion}
\subsection{Methodological Interpretation}
The present pilot study was designed to establish a defensible, field-specific pathway from SPHEREx diffuse-background estimation to a first dust-opacity diagnostic. The central methodological result is that the dust analysis must begin with an explicit assessment of channel support. In the current $l\simeq300^\circ$ field analysis, that check is straightforward: all 102 mosaicked channels have uniformly high coverage, so the long-wavelength restriction used below is a science choice rather than a support-driven cut.

That point sharpens the interpretation of the pilot. The current limitation is not the absence of spectral sampling; SPHEREx already provides a densely sampled channel set over most of the 0.75--5.0~$\mu$m range \citep{korngut2018,crill20,bock26}. In the present analysis, the detector-level products already provide uniformly high support across the full channel set, and the current pilot extinction analysis can therefore be carried out over the cleaner 3.205--4.98~$\mu$m interval as a science choice rather than as a support-driven necessity.

The second conclusion is methodological. Large-scale background estimation remains a source of systematic uncertainty, exactly as earlier mid-infrared extinction studies had already implied \citep{butler09,butler12,lim14,lim15,lim16}. The direct comparison against the earlier masked-median background model shows that modest channel-by-channel background differences can propagate into substantially larger differences in the normalized extinction curve at short wavelengths. The correct reading of the current model comparison is therefore not that a unique grain model has been identified, but that SPHEREx now makes it possible to quantify the sensitivity to the background prescription while still recovering a field-specific relative-opacity proxy from the local data alone. This makes the LMF result a useful reference curve for testing which dust-model directions are favored, while keeping the background dependence explicit.

\subsection{Dust-model Comparison}
Figure~\ref{fig:spitzercurve} and Table~\ref{tab:dustmodels} summarize the model assumptions needed to interpret the opacity comparison. Within the OH94 family, the main transition sampled by the pilot curve is not a change in the underlying silicate versus carbon core chemistry, but a change in mantle growth and the degree of dense-core coagulation applied to the same basic grain population. The residuals in Section~3.3 describe shape agreement under the adopted normalization, rather than a measurement of mantle thickness. The O11 sequence probes a related but not identical axis: it follows the time-dependent opacity evolution of ice-coated silicate and graphite aggregates, so agreement with the earlier O11 snapshots is read mainly as sensitivity to the onset of aggregate growth and porosity evolution rather than simply to a larger mantle on otherwise compact grains. The D22 comparison set probes a different axis again: the E MRN, E tau1, and E tau3 branches are laboratory-informed grain-growth realizations built from ellipsoidal silicate grains coated by H$_2$O- and CO$_2$-rich ice mantles, so their main leverage is on how grain-growth and ice-profile distortion reshape the mid-infrared opacity curve within the D22 framework. The bottom-panel OpTool realizations then separate those effects in a deliberately controlled way by isolating grain-size and ice-fraction changes within the adopted optical formalism. What they show more robustly is that the inferred ranking is highly sensitive to the treatment of local spectral structure, normalization, and background systematics.

This interpretation remains qualitative because the comparison is based on relative shape agreement after normalization rather than on a formal likelihood-based fit with free physical parameters. The current figures do not yet justify quoting a unique density, a unique coagulation time, or a unique ice mass fraction for SDC300.257$-$0.253. The SPHEREx-only and combined tests differ in wavelength coverage and normalization (4 and 24~$\mu$m), so their residual orderings need not agree. In other words, the fitting result is already methodologically informative, but the interpretation should be that model discrimination remains highly sensitive to the treatment of local spectral structure and background definition, not that one exact dust prescription has been uniquely measured.

\subsection{Local Spectral Structure at 3--5~$\mu$m}
The same caution applies to the local structure seen near 3.3 and 4.2~$\mu$m in the pilot curve. Backtracking those apparent bumps to the actual SPHEREx channel grid shows that the strongest local maxima fall in the channels centered at 3.205~$\mu$m (3.16--3.25~$\mu$m) and 4.250~$\mu$m (4.23--4.27~$\mu$m), with weaker neighboring structure around 4.03 and 4.14~$\mu$m. Those enhancements are astrophysically suggestive, but at the present stage they should be interpreted as features of the effective extinction proxy rather than as secure detections of narrow opacity bands. A bump near 3.3~$\mu$m could reflect real hydrocarbon-sensitive structure, but it is also the wavelength regime in which PAH-sensitive background emission can imprint residual spectral curvature if the reconstructed background is not perfectly matched to the true diffuse field behind the source \citep{allamandola89,peeters02,smith07,tielens21}. Likewise, the enhancement near 4.2~$\mu$m is qualitatively consistent with the spectral neighborhood of the 4.27~$\mu$m CO$_2$ ice band, yet the current SPHEREx result alone does not justify treating that correspondence as a direct identification \citep{pontoppidan08,boogert15}. Because the plotted quantity is derived from $I_{\rm obs}/I_{\rm bg}$, any wavelength-dependent mismatch in the background estimate or in the local continuum can create or amplify apparent bumps in the normalized relative-opacity proxy. These features are therefore best regarded as motivated targets for the next round of diagnostics, in which the source-aperture spectra of $I_{\rm obs}(\lambda)$, $I_{\rm bg}(\lambda)$, and their ratio will be inspected explicitly before assigning them to specific grain or ice carriers.

\subsection{External Comparison and Survey Outlook}
The external archival comparison provides an independent check on the same background-dependent opacity proxy. Once the archival maps are kept on their native grids and evaluated through the SDC300.257$-$0.253 aperture, the LMF background yields a comparatively stable set of IRAC, WISE W3 and W4, MIPS 24~$\mu$m, and Herschel 70~$\mu$m points. The key additional step is that the WISE surface-brightness scale is tied to Spitzer through background-subtracted empirical regressions instead of being taken directly from the raw atlas calibration. Normalizing the full relative-opacity comparison at 24~$\mu$m is useful because it ties the empirical curve and the theoretical models to the same mid-infrared anchor actually measured in the field. On that scale, the recalibrated WISE points, the MIPS 24~$\mu$m point, and the Herschel 70~$\mu$m point define a coherent external sequence for the adopted source aperture. By contrast, the masked-background model remains much less stable and can drive several of the external proxies negative. That behavior does not by itself prove that the LMF solution is unique, but it does reinforce the decision to use the LMF as the main background definition for the current pilot. It also removes an unnecessary reprojection step from the external comparison, which is methodologically cleaner because the background is estimated from the native archival surface-brightness structure rather than from a grid resampled to match SPHEREx.

More generally, the relative-opacity curve assembled from SPHEREx, IRAC, WISE W3 and W4, MIPS, and Herschel should not yet be read as a fully foreground- and background-decomposed extinction law. Each band still carries its own beam, calibration convention, large-scale zero level, and local background and foreground ambiguity, so systematic differences between bands remain unavoidable at the present stage \citep{butler09,butler12,battersby11,lim14,lim15,lim16}. The value of the current multiband sequence is therefore diagnostic rather than definitive: it identifies which spectral structures persist once a common LMF background methodology is imposed, and it shows that the present band-by-band backgrounds are already stable enough to motivate targeted follow-up tests of specific wavelengths, source apertures, and background conventions. In that sense the present analysis sits naturally alongside modern dust studies that combine extinction, thermal emission, and grain-evolution arguments across multiple environments rather than treating any one tracer as complete by itself \citep{jones13,fanciullo15,kohler15,hensley21}.

A related caveat concerns the adopted dark source itself. The \citet{peretto09} and \citet{peretto16} IRDC catalogs were designed to identify infrared-dark structures efficiently over the Galactic plane, not to guarantee that every cataloged source is maximally dark or geometrically ideal for extinction-law work. SDC300.257$-$0.253 is therefore useful as a pilot aperture, but it may not be the optimal object for a final opacity analysis. Future source-by-source work should revisit the depth, isolation, and background contrast of the adopted structures before treating the resulting curve as a definitive extinction law. At the same time, that caution does not weaken the scientific motivation for IRDC-focused follow-up. If such clouds do trace early stages of massive star and star-cluster formation \citep{mckee03,tan14}, then a survey-scale background framework that can be carried across many IRDC environments is valuable precisely because it allows the same opacity methodology to be tested over a broader population rather than on a single hand-tuned source. In that sense, the present LMF background is valuable because it already provides a stable and operationally uniform background definition across the supported channel set, which is exactly the property needed for diagnostic follow-up and for survey-wide background products built on a common convention.

The broader significance of the pilot follows from the mission delivery cadence. SPHEREx was designed as a repeated all-sky spectral survey with public spectral-image products and survey-scale science applications \citep{korngut2018,crill20,crill25,bock26}. The planned annual public releases will provide reprocessed calibrated spectral images and higher-level survey products \citep{bock26}. For the problem considered here, these releases matter because they should expand the number of fields for which similarly high, near-uniform channel support can be reached. As the quick-release products continue to be recalibrated and the mission advances toward the eventual Year~2 release over the nominal 2~yr survey \citep{bock26}, the support uniformity, calibration stability, and background fidelity should improve further. The main implication is that the same LMF framework can move from a field-specific pilot to a survey-scale baseline: a uniform set of diffuse-background products can be generated first, and cloud-specific refinements can then be measured against that common reference. Such a baseline would be useful well beyond the specific opacity curves shown here, because it would enable direct comparisons to near-infrared extinction maps built with NICE, NICER, and NICEST \citep{lada94,lombardi01,lombardi09}, to mid-infrared mass--surface density analyses \citep{butler09,butler12,battersby11,lim14,lim15,lim16}, and to broader questions of dust evolution and infrared mass estimation.

This background-estimation framework extends naturally to the Probe Far-infrared Mission for Astrophysics (PRIMA). Like SPHEREx, its imager PRIMAger will use linearly variable filters (LVFs), sampling 24--84~$\mu$m to fill gaps between legacy broadband measurements and strengthen constraints on dust emission and grain evolution \citep{crill20,ciesla25}. This approach offers compelling potential by linking SPHEREx extinction diagnostics with PRIMA dust-emission studies, further enriched by complementary stellar-extinction measurements from the Nancy Grace Roman Space Telescope \citep{akeson19}.

\section{Summary}
We have presented a SPHEREx diffuse-background analysis in a Galactic-plane field centered at $(l,b)=(300^\circ,0^\circ)$ using the QR2 detector-level products. In the present analysis, the full 102-channel mosaic set has uniformly high coverage, yielding a channel-resolved background set for the field itself.

We then used the 14~arcmin LMF method to construct diffuse-background maps for all 102 supported channels and carried the extinction analysis forward only for the cleaner long-wavelength subset of 41 channels spanning 3.205--4.98~$\mu$m. For the IRDC SDC300.257$-$0.253, this yields a field-specific SPHEREx extinction and relative-opacity proxy built directly from local data products. The main scientific result is methodological: changing the background prescription substantially changes the derived 3--5~$\mu$m curve. In the present field, comparison with the earlier masked-median background model gives a median channel-by-channel background difference of about 6\% and a maximum change of $|\Delta\tau_{\rm norm}|\simeq 7.4$ near 3.205~$\mu$m. An empirical empty-aperture analysis provides a compact diagnostic of the residual fluctuation floor left by the channel-by-channel background reconstruction.

The same framework can already be extended beyond SPHEREx. Using the identical source aperture on native-grid IRAC, WISE W3 and W4, MIPS, and Herschel maps, we obtain a same-source comparison from 3.6 to 70~$\mu$m once the WISE residual maps are empirically tied to Spitzer after background subtraction. The model comparisons illustrate the background framework but do not identify a unique dust population for this source. The significance of the present study is therefore clear: SPHEREx already supports a practical survey-baseline method for channel-resolved diffuse-background modeling in the Galactic plane, while also showing explicitly how strongly dust-model interpretation depends on the adopted background and on the treatment of localized spectral structure. As later SPHEREx public releases improve support uniformity and calibration stability \citep{bock26}, the same LMF framework can be scaled from this field-specific demonstration to survey-wide background products and then used as the common reference for extinction, mass--surface density, and dust-evolution studies across the Galactic plane.

\begin{acknowledgments}
This work is based on preliminary SPHEREx QR2 analysis for diffuse Galactic background studies. W.L. is supported by Caltech/IPAC under contract No. 80GSFC21R0032 with the National Aeronautics and Space Administration. Y.J. is supported by National Research Foundation of Korea (NRF) grants funded by the Korea government (MSIT; Nos. 2021R1A2C1011718 and RS-2024-00416859).
\end{acknowledgments}

\end{document}